\documentclass{iopart}
\usepackage{times}
\begin{document}

\title{Angular Momentum and the Formation of Stars and Black Holes}

\author{Richard B Larson}

\address{Department of Astronomy, Yale University,
         Box 208101, New Haven, CT 06520-8101, USA}

\ead{richard.larson@yale.edu}

\begin{abstract}

   The formation of compact objects like stars and black holes is strongly constrained by the requirement that nearly all of the initial angular momentum of the diffuse material from which they form must be removed or redistributed during the formation process.  The mechanisms that may be involved and their implications are discussed for (1) low-mass stars, most of which probably form in binary or multiple systems; (2) massive stars, which typically form in clusters; and (3) supermassive black holes that form in galactic nuclei.  It is suggested that in all cases, gravitational interactions with other stars or mass concentrations in a forming system play an important role in redistributing angular momentum and thereby enabling the formation of a compact object.  If this is true, the formation of stars and black holes must be a more complex, dynamic, and chaotic process than in standard models.  The gravitational interactions that redistribute angular momentum tend to couple the mass of a forming object to the mass of the system, and this may have important implications for mass ratios in binaries, the upper stellar IMF in clusters, and the masses of supermassive black holes in galaxies.

\end{abstract}

\section{Introduction}  

   It has long been recognized that the biggest obstacle to the formation of stars from diffuse gas is that a star can contain only a tiny fraction of the initial angular momentum of the gas from which it forms, so that nearly all of this angular momentum must be removed or redistributed during the formation process (Mestel 1965; Spitzer 1968, 1978; Bodenheimer 1995; Larson 2003b; Jappsen and Klessen 2004).  The specific angular momenta of typical star-forming molecular cloud cores are at least three orders of magnitude larger than the maximum specific angular momentum that can be contained in a single star, even when rotating at breakup speed (Bodenheimer 1995).  The `angular momentum problem' is a familiar one for stars like the Sun, but it may be more severe for massive stars whose matter must come from a larger region containing more angular momentum, and it may be especially so for black holes in galactic nuclei because they are smaller than stars in relation to the size of the system in which they form.  The masses that massive stars and central black holes can attain may therefore be limited by the efficiency with which angular momentum can be removed during the formation process.

   The angular momentum problem was first studied in the context of single stars forming in isolation (Mestel 1965; Spitzer 1968), but it now seems likely that most stars form not in isolation but in systems such as binary or multiple systems or clusters, and in this case, it is necessary to consider both the orbital and the spin components of the angular momentum of the matter from which each star forms.  If a star-forming cloud core forms a binary or multiple system, some of its angular momentum evidently goes into stellar orbital motions, plausibly accounting for the orbital component of the angular momentum of the matter from which each star forms, but the spin angular momentum of this matter must still be removed or redistributed during the star formation process.  The excess spin angular momentum of the matter from which each star forms could in principle be transferred to outlying gas or to the orbital motions of other stars, and both magnetic and gravitational forces can play important roles in this loss or redistribution of angular momentum.

   As will be reviewed in Section 2, magnetic torques can remove angular momentum from diffuse star-forming clouds and from accreting protostars during the final stages of accretion, but this leaves a large intermediate range of densities where magnetic coupling is weak and gravity dominates the dynamics, governing not only the collapse of a molecular cloud or cloud core but also the redistribution of angular momentum within it.  Simulations of star formation show the appearance of trailing spiral features within which gravitational torques can transport angular momentum outward, and forming stars can also lose orbital energy and angular momentum to the surrounding gas by gravitational drag, causing them to spiral together and form more compact systems.  Within these compact systems, tidal torques between forming stars and the gas orbiting around other forming stars can transfer the angular momentum of this gas to stellar orbital motions, allowing the gas to be accreted by the forming stars.

   If companion stars in binary or multiple systems or clusters play an important role in absorbing and redistributing the excess angular momentum of forming stars, few stars may form in complete isolation.  The formation of massive stars in clusters and of central black holes in galaxies may be even more dependent on the presence of a surrounding system to absorb and redistribute the larger amount of angular momentum involved.  The fact that the mass of the most massive star in a cluster and the mass of the central black hole in a galaxy both increase systematically with the mass of the surrounding system suggests that the associated system does indeed play an essential role in the formation of these objects.

   If the solution of the angular momentum problem partly involves gravitational interactions with other stars in an associated system, star formation must then be a more violent and chaotic and variable process than in standard models for isolated star formation.  Chaotic formation processes are hinted at by the fact that the spin axes of both stars in clusters and central black holes in galaxies are randomly oriented and not correlated with the properties of the surrounding system, suggesting that chaotic interactions randomize the residual angular momentum of forming objects during the formation process.

\section{Transport processes in collapsing clouds}  

   Magnetic and gravitational forces can both play important roles in transporting angular momentum in star-forming clouds and solving the angular momentum problem.  The early low-density stages of molecular cloud evolution may be magnetically dominated because the degree of ionization is high enough that the gas is strongly coupled to a magnetic field, and with typical observed field strengths, magnetic forces can then exceed thermal pressure and have important effects on cloud evolution (Heiles \etal 1993; McKee \etal 1993; Crutcher 1999).  Magnetic braking, in particular, may remove much of the initial angular momentum of a star-forming cloud core during the early stages of its evolution, transferring it to diffuse surrounding gas, and this may determine the amount of angular momentum remaining in the dense regions that eventually collapse to form stars (Mouschovias 1977, 1991; Basu and Mouschovias 1994).  Most prestellar cloud cores are observed to rotate considerably more slowly than would be expected if they had condensed from diffuse interstellar matter with no loss of angular momentum (Goodman \etal 1993), and this slow rotation can plausibly be attributed to the effect of magnetic braking during the early stages of cloud evolution. 

   The specific angular momenta of the molecular cloud cores studied by Goodman \etal (1993), Barranco and Goodman (1998), Jijina \etal (1999), and Caselli \etal (2002) are comparable to those of wide binary systems with separations of a few thousand AU, and this suggests that magnetic braking may already have removed enough angular momentum from these cores for them to form wide binaries without further loss of angular momentum (Mouschovias 1977; Bodenheimer 1995; Simon \etal 1995; Larson 2001, 2003b).  However, typical binaries with a median separation of only $\sim 30$ AU (Duquennoy and Mayor 1991) have specific angular momenta about an order of magnitude smaller than those of the observed cloud cores, so their formation requires some further loss of angular momentum beyond that provided by magnetic braking during the early stages of cloud evolution (Larson 2001, 2003b; Fisher 2004).  In addition, even in wide binaries, the spin angular momentum of the matter from which each individual star forms must somehow be removed or redistributed during the formation process.

   As a cloud core contracts, magnetic flux is gradually lost by ambipolar diffusion, and magnetic effects therefore become progressively less important while gravity becomes increasingly dominant (Basu and Mouschovias 1994; Mouschovias and Ciolek 1999; Nakano \etal 2002).  The observed star-forming cloud cores are roughly magnetically critical, having magnetic support comparable to gravity, so they may represent a transition stage between regimes of magnetic dominance and gravitational dominance.  Because magnetic braking eventually becomes unimportant, angular momentum is predicted to be nearly conserved during the later stages of collapse (Basu and Mouschovias 1994, 1995; Basu 1997), and observations suggest that angular momentum is indeed approximately conserved in regions smaller than about 0.03 pc (Ohashi \etal 1997; Ohashi 1999; Myers \etal 2000; Belloche \etal 2002).  Eventually most of the initial magnetic flux is lost, leaving a residual field of only about 0.1 G that is comparable to the magnetic field strength in the early Solar System inferred from meteorites but too small to be dynamically important (Tassis and Mouschovias 2007).

   Once gravity gains the upper hand, runaway collapse toward a density singularity always occurs and proceeds qualitatively as in the spherical case even if rotation and magnetic fields are present, as long as these effects are not strong enough to completely prevent collapse (Larson 2003b, 2007).  Calculations of axisymmetric collapse with rotation show that a central density singularity develops even if the angular momentum of each fluid element is conserved (Norman \etal 1980; Narita \etal 1984; Matsumoto \etal 1997).  However, this central density singularity can develop into a star only if most of the angular momentum of the gas orbiting around it is removed.  If there are any departures from axial symmetry in this surrounding gas, gravitational torques will be present, and these torques will transport angular momentum outward in the presence of trailing spiral density fluctuations like those that occur ubiquitously in simulations of star formation (Larson 1984).  Jappsen and Klessen (2004) find that the specific angular momentum of the gas in a collapsing cloud core can be reduced by an order of magnitude in this way, and they also find that the resulting distribution of specific angular momenta is compatible with the distribution of specific angular momenta of observed binary systems (Duquennoy and Mayor 1991).

   After stars begin to form, gravitational redistribution of angular momentum can continue to be important because the stars continue to interact strongly with the remaining gas.  For example, in a forming binary or multiple system, the stars can lose orbital energy and angular momentum to the surrounding gas by gravitational drag or `dynamical friction' effects, causing them to spiral together and form more tightly bound systems (Larson 1990a, 2001).  In these more tightly bound systems, the stars can then exert decelerating tidal torques on the residual gas orbiting around other forming stars, transferring its angular momentum to stellar orbital motions and allowing it to be accreted by the growing stars (Larson 2002; see Section 3.2 below.)  If a residual magnetic field is important it may tend to inhibit such fragmentation and gravitational transport effects, but even in the limiting case where there is no ambipolar diffusion and the field remains fully coupled to the gas, simulations have shown that a magnetic field cannot prevent binary formation (Price and Bate 2007; Hennebelle and Teyssier 2008).

   Eventually the gas near a forming star can become ionized and recouple to the magnetic field, twisting and amplifying it (Machida \etal 2007).  Stellar dynamo activity may also contribute to the formation of a magnetospheric region around the star, and gas that falls into this region may be captured and fall onto the star along magnetic field lines, as in current models of accretion onto T Tauri stars (Hartmann 1998; Romanova \etal 2008).  Rapidly spinning magnetospheric regions may also generate the bipolar jets characteristic of newly formed low-mass stars, and most of the angular momentum of the gas that enters such a region may be carried away by a rotating jet or wind (Tomisaka 2002; Banerjee and Pudritz 2006; Ray \etal 2007; Machida \etal 2007).  A rotating wind may also remove angular momentum from the innermost partially ionized part of a protostellar disk (Shu \etal 2000; Pudritz \etal 2007).  Rotating outflows have been observed and can carry significant angular momentum per unit mass (Coffey \etal 2007, 2008), but the objects studied so far have mass loss rates that are too small for these outflows to remove a large amount of angular momentum from a forming star.  If a rotating magnetosphere or magnetized inner disk with a radius of several stellar radii were to accrete all the gas that falls into it and expel all of its angular momentum in a rotating outflow, the angular momentum problem would be somewhat reduced, but the angular momentum of the gas in a typical star-forming cloud core would still have to be reduced by more than two orders of magnitude by other effects like those discussed above for it to fall into this region and be accreted.

\section{Formation of low-mass stars}  

   The angular momentum problem has been studied most extensively in connection with the formation of low-mass stars like the Sun.  Much evidence suggests that these stars form in dense cloud cores that have masses similar to those of the stars that form in them, but specific angular momenta that are at least three orders of magnitude larger than the maximum that can be contained in a single star (Bodenheimer 1995) and two orders of magnitude larger than can be contained even in a rapidly rotating magnetospheric region.  More than 99\% of the initial angular momentum of the gas from which each star forms must therefore be removed during the formation process.  Where does this angular momentum go, and how is it transported?  In standard models for isolated star formation (Shu \etal 1987, 1993, 1999), most of the angular momentum of a rotating cloud core goes into a circumstellar disk and is transported outward in this disk by an assumed viscosity.  However, for most stars the formation process must be much more complicated than this because most stars form not in isolation but in binary or multiple systems where the companions can have major effects on the redistribution of angular momentum.  Both disk processes and interactions with companions therefore need to be considered.

\subsection{Disk processes}  

   In standard models for isolated star formation, stars gain most of their mass from a circumstellar accretion disk in which angular momentum is transported outward by some assumed disk viscosity.  If all of the angular momentum of a typical star-forming cloud core is to end up in a circumstellar disk, however, a very large disk is required, possibly as large as the cloud core itself (Larson 2002).  Circumstellar disks are often observed around newly formed stars and are also frequently found in simulations of star formation (e.g., Bate \etal 2003), but most of the observed and simulated disks are much too small in size and mass to account for more than a small fraction of the angular momentum of a typical star-forming cloud core, and this is true also of the `Solar Nebula' from which our Solar System is believed to have formed (Larson 2002).  Circumstellar disks therefore cannot by themselves account for more than a small fraction of the angular momentum of star-forming cloud cores.  Disks could nevertheless play an important role in transporting angular momentum during early stages of the star formation process, and both gravitational torques in sufficiently massive disks and magnetic torques in sufficiently ionized disks might be important (for reviews, see Larson 1989, 2003; Hartmann 1998; Stone \etal 2000; Lodato 2008).

   A general limitation of disks as a transport medium, however, is that they are fragile structures that cannot support a large torque stress; this is because, for most of the proposed transport mechanisms, the torque stress is limited by the thermal pressure in the disk and typically saturates at a level corresponding to a value significantly less than unity for the `alpha' parameter of Shakura and Sunyaev (1973) giving the ratio of torque stress to thermal pressure.  This implies long timescales for angular momentum transport that are typically thousands to hundreds of thousands of orbital periods (Larson 1989, 2002).  Such long timescales are problematic because most circumstellar disks may not survive this long before being disrupted by interactions with other stars in a complex environment; for example, in the simulations of the formation of clusters of stars by Bate \etal (2003), Bate and Bonnell (2005), and Bate (2009a), disks are constantly formed and destroyed but few survive for the entire duration of the simulation.  Bate (2009a) estimates that the median truncation radius of disks due to encounters is only about 3 AU, but notes that disks may grow back to larger sizes after truncation.  Large long-lived disks like those postulated in standard models of star formation are found only around stars that have been ejected from the system and whose formation process has already been largely completed.  Such disks may therefore be rare, or may occur only in the final stages of the formation of relatively isolated stars.  More compact circumstellar disks may, however, be more common and may play an important role in the formation of binary and multiple systems where tidal interactions can transfer angular momentum from circumstellar disks to the stellar orbital motions (Bate and Bonnell 1997; Bate 2000; Blondin 2000; Boffin 2001; Larson 2002; see below).

   If a circumstellar disk becomes too massive, it may become gravitationally unstable and fragment into clumps and filaments.  One possible outcome is that temporarily bound clumps may form and lose enough angular momentum to the rest of the disk to spiral inward and fall onto the central object (Vorobyov and Basu 2006).  The resulting bursts of accretion may contribute significantly to the growth in mass of the central object and may help to account for the variability of newly formed stars, but they do not change the overall timescale for protostellar accretion.  Another possibility is that one or more bound companions such as a massive planet or a small companion star may form in the disk, as often occurs in simulations (Larson 1978; Bonnell and Bate 1994; Boss 2002; Bate \etal 2003; Bate and Bonnell 2005).  If a massive planet forms, the torques exerted by the planet on the disk can transport angular momentum from the inner to the outer part of the disk and may thus help to drive an inflow toward the central star; a Jupiter-mass planet can be as important in this respect as transient spiral density fluctuations, and planets larger than Jupiter can be much more important because the tidal torque increases with the square of the planet's mass (Larson 1989).  If a binary companion forms, its effects may even completely dominate the subsequent evolution of the system, as will be discussed in Section 3.2.

   In addition to gravitational and magnetic effects, pressure effects can also play a role in disks if acoustic waves are generated in them by external disturbances.  Acoustic waves carry both energy and momentum, and in a disk they carry angular momentum; if these waves take a trailing spiral form because of differential rotation, then they always transport angular momentum outward, regardless of their direction of propagation (Larson 1989).  For a given spiral pattern, acoustic transport of angular momentum exceeds gravitational transport if the disk is gravitationally stable and has Toomre stability parameter $Q > 1$ (Larson 1989).  Pfalzner and Olczak (2007) have shown that gravitational transport of angular momentum in disks can be triggered by tidal interactions with companions, and since acoustic transport can be equally or more important if a disk is stable, tidal interactions may quite generally induce outward transport of angular momentum in disks.

\subsection{The role of binary and multiple systems}  

   The available evidence suggests that most stars form in binary or multiple systems.  About half of all solar-type stars have fainter binary companions, and the fraction with companions increases with stellar mass, just as would be expected if most stars form in unstable multiple systems that preferentially eject the less massive stars and retain the more massive ones in binaries (Heintz 1978; Abt 1983; Duquennoy and Mayor 1991; Mathieu 1994; Zinnecker and Mathieu 2001; Goodwin \etal 2007).  Lada (2006) has noted that because the binary frequency is less than half for the M stars, the most common class of stars, the overall observed binary frequency is less than half; however, the fraction of stars formed in binary or multiple systems may be considerably higher than the fraction presently observed in such systems because many systems may be disrupted by interactions soon after formation.  This possibility is supported by the fact that the binary fraction is typically higher in star-forming clouds than in the field (Duch\^ene \etal 2007).

   The statistics of single and binary systems can be roughly accounted for if most stars form in triple systems that decay into similar numbers of binary and single stars (Larson 1972; Reipurth 2000; Goodwin and Kroupa 2005).  Most of the youngest known stars, the embedded jet sources, appear to be in triple systems (Reipurth 2000, 2001), and it has been suggested that interactions in such systems may play a role in jet production by perturbing circumstellar disks (Reipurth 2001).  Because the least massive stars in such systems are preferentially ejected while the more massive ones are retained in binaries, the resulting binary frequency should increase with mass, as is indeed observed: the binary fraction is $\sim 30$\% among M stars (Fischer and Marcy 1992), $\sim 50$\% among G stars (Duquennoy and Mayor 1991), and 70\% or more among O stars (Preibisch \etal 2001; Mason \etal 2009).  A similar dependence of binary frequency on mass has been predicted theoretically in a large simulation of cluster formation by Bate (2009a).

   Binary and multiple systems also appear frequently in simulations of the formation of groups and clusters, regardless of the techniques or assumptions used in the simulations (Larson 1978; Bodenheimer \etal 1993, 2000; Zinnecker and Mathieu 2001; Bate \etal 2002, 2003; Bate and Bonnell 2005; Bate 2009a,b).  Both observations and simulations therefore indicate that binary and multiple systems are a common outcome of star formation that depends only on very general features of the dynamics of collapsing clouds.  The simulations typically show that the most massive star in a binary forms first, while the less massive one forms shortly afterward from gas that has too much angular momentum to join it (see also Attwood \etal 2009).  In such cases nature may solve the angular momentum problem by putting the excess angular momentum into a binary companion.

   Most of the multiple systems formed in the simulations soon decay and eject stars, sometimes at high speed (Bate \etal 2002, 2003; Delgado-Donate \etal 2004; Bate and Bonnell 2005; Bate 2009a).  The escapers may then carry away angular momentum from these systems that eventually ends up in the random motions of field stars.  Many single stars like the Sun could plausibly have originated in this way, in which case they would have had a more chaotic and violent early history than in standard models.  Evidence that our Solar System experienced a disturbance in its early history is provided by the fact that its fundamental plane is tilted 8 degrees from the solar rotation axis, plausibly because of an early encounter with another star (Heller 1993).  Such encounters might have played a role in shaping the properties of our Solar System, and they might also account for the properties of the many extra-solar planetary systems that have massive planets in close eccentric orbits (Malmberg and Davies 2009).

   Several gravitational effects can contribute to the redistribution of angular momentum in a forming stellar system.  The forming stars can at an early stage lose orbital energy and angular momentum to the surrounding gas by gravitational drag, causing them to spiral together and form more tightly bound systems.  This effect is important in simulations where it can result in the formation of close binary systems from stars that initially formed at much larger separations (Bate \etal 2002; Bate and Bonnell 2005; Bate 2009a,b).  Interactions between stars and circumstellar gas around other stars can also dissipate orbital energy and lead to the formation of more tightly bound systems (Larson 1990a; Heller 1995; Bate \etal 2002; Bate and Bonnell 2005; Bate 2009a).  Tidal interactions in these more compact systems can then transfer angular momentum from the matter orbiting around the forming stars to the orbital motions of companion stars (Larson 2002).  This occurs, for example, in simulations of binary formation where both stars have circumstellar disks that are tidally perturbed (Bate 2000; Krumholz \etal 2009); the tidal disturbance creates a trailing density enhancement in each disk, and the gravitational torque between this trailing response and the companion star transfers angular momentum from the disk to the companion.  Through such a combination of gravitational drag and tidal effects, the excess angular momentum of he gas from which each star forms can be transferred partly to the orbital motions of other stars in the system and partly to the surrounding gas.

  If the matter in a perturbed disk is to be accreted by the central star, angular momentum must still be transported outward in the disk, but the problem of long transport times is here alleviated by the fact that the disk is tidally truncated, so that angular momentum does not have to be transported very far.  Several mechanisms may contribute to the transport of angular momentum in tidally disturbed disks, including both gravitational and wave torques.  Waves of any kind that are induced by an external tidal disturbance, including acoustic waves, density waves in which gravity plays a role, and MHD waves, all carry negative angular momentum and therefore can generate an inflow if there is any dissipation of these waves, for example by shocks (Larson 1989, 1990b; Blondin 2000).  In the simulations of Bate (2000) and Krumholz \etal (2009), some combination of these effects plus numerical viscosity transfers most of the angular momentum of a forming binary system to the stellar orbital motions, allowing the stars to accrete most of the gas present.

   In an eccentric binary, periodic close encounters between a companion and a circumstellar disk may trigger bursts of accretion onto a forming star, perhaps helping to explain flareups such as the FU Orionis phenomenon (Bonnell and Bastien 1992).  In a cluster of forming stars, encounters with passing stars can also remove angular momentum from circumstellar disks and thus help to drive inflows (Ostriker 1994; Pfalzner \etal 2008).  Even if the gas around a forming star is not in a disk, interactions with passing stars can remove angular momentum from it and cause some of it to be accreted, as sometimes happens in simulations (e.g., Bate \etal 2003; Bate and Bonnell 2005).  Interactions with other stars may therefore tend quite generally to remove angular momentum from the gas around forming stars.  The angular momentum removed is transferred to the orbital motions of the other stars, and much of it may therefore ultimately end up in the random motions of field stars.

\subsection{Implications for binary mass ratios}  

   In a forming binary system, the mass acquired by each star depends on the amount of angular momentum transferred to the companion or to surrounding gas.  A binary system with a given mass and angular momentum can contain two stars of equal mass in a relatively close orbit, or two stars of unequal mass in a wider orbit.  In a system with unequal masses, the more massive star may accrete less rapidly because of the weak tidal effect of the relatively small and distant companion, while the less massive companion may accrete more readily from a circumbinary disk (Artymowicz 1983; Bate 1997, 2000).  Such effects may favor the growth of the less massive star at the expense of the more massive one, causing the masses to equalize.  Equal masses may also be favored by the fact that the total energy of a binary system with a given total mass and angular momentum is minimized for stars of equal mass in circular orbits.  Simulations of binary formation do, in fact, show a tendency for the stars in close binaries to acquire similar masses (Bate 2000; Bate \etal 2002; Bate 2009a; Krumholz \etal 2009).

   Observations also show that the mass ratios of the stars in binary systems tend to be closer to unity than would be predicted if their masses were selected randomly from a standard IMF.  This is especially true for close binaries, whose distribution of mass ratios of is roughly flat in linear units; for wider binaries, this distribution generally declines with increasing mass ratio, but never as steeply as would be predicted for random selection from a standard IMF (Abt 1983; Trimble 1990; Mayor \etal 2001; Goldberg \etal 2003; Mazeh \etal 2003).  The masses of the stars in binary systems are therefore correlated, especially for close binaries, and this can be understood if tidal interactions play an important role in the formation of these systems, causing the growth rate of each star to depend on the mass of the companion and increase with the companion's mass.

\subsection{Summary}  

   Summarizing the conclusions of this section, observations and simulations both suggest that most stars form in binary or multiple systems whose orbital motions account for at least part of the initial angular momentum of star-forming cloud cores.  Gas drag effects can make these systems more compact, and tidal interactions in the resulting compact systems can transfer angular momentum from the gas around each forming star to the orbital motions of the companions.  In close binaries, this tidal coupling tends to equalize the stellar masses.  Companion stars may thus often play an essential role in absorbing and redistributing the excess angular momentum of forming stars.

\section{The formation of massive stars}  

   The formation of massive stars is less well understood than that of low-mass stars, and two hypotheses that have been much debated are (1) massive stars form by the direct collapse of massive cloud cores, and (2) they form along with other stars by accretion in dense cluster cores (Zinnecker and Yorke 2007; Krumholz and Bonnell 2009).  In either case, the angular momentum problem remains, and it may be even more severe than for low-mass stars because more mass and angular momentum need to be redistributed.  If massive stars typically form in clusters, as observations suggest, the matter from which each massive star forms may come from a region containing many other forming stars, and the assembly of this matter into a single star may then be a complex process involving interactions with many other stars.

   The role of magnetic fields is less clear for massive stars than for low-mass stars because massive star-forming clouds have higher mass-to-flux ratios than less massive clouds, making magnetic fields less important in relation to gravity; in addition, massive young stars lack the highly collimated jets characteristic of low-mass stars (Arce \etal 2007), so there is no similar evidence that magnetic fields control the final stages of star formation.  The role of disks is also less clear because there is less evidence for circumstellar disks around massive young stars.  Extended flattened or toroidal structures are sometimes seen around massive newly formed stars (Cesaroni \etal 2007; Beuther \etal 2007), but these structures are thousands of AU in radius and their role in the formation of the central object is unclear.  The paucity of evidence for circumstellar disks around massive newly formed stars may partly reflect observational limitations, but it may also reflect the fact that massive stars form in complex and crowded environments where it is difficult for disks to form and survive.

\subsection{The role of multiple systems and clusters}  

   There is, on the other hand, abundant evidence that nearly all massive stars form in binary and multiple systems and clusters (Abt \etal 1990; Kobulnicky and Fryer 2007; Zinnecker and Yorke 2007).  In the well-studied young cluster NGC 6231, for example, the fraction of O stars that are in binaries is at least 70\% and may approach 100\% (Sana \etal 2008).  Massive stars also often occur in Trapezium-like multiple systems, many of whose members are themselves binary or multiple; the Orion Trapezium, for example, contains at least 11 stars (Zinnecker and Yorke 2007).  Many close O-star binaries may therefore originate from the decay of Trapezium-like multiple systems.  Because the decay of multiple systems preferentially ejects the less massive stars and retains the more massive ones in binaries, the resulting binary fraction should increase with mass, as is indeed observed (see Section 3.2).

   The companions of massive stars are themselves typically massive and close; for example, more than 70\% of the O stars in NGC 6231 are in O+O or O+B binaries, and half of these are close binaries (Sana \etal 2008).  Massive stars therefore typically form near other massive stars, and interactions with these massive companions may play an important role in their formation.  In addition, massive stars are also strongly concentrated in clusters and associations and are often found in Trapezium-like multiple systems in dense cluster cores.  Even most of the minority of O stars that appear isolated can be explained as escapers from clusters or associations, leaving only a small fraction of O stars that cannot clearly be traced back to such an origin, estimated as 4 percent by de Wit \etal (2005) and at most 5 to 10 percent by Zinnecker and Yorke (2007).  Thus, massive stars typically form with more massive companions and in larger systems than low-mass stars.

   All of these observed features of massive star formation are found in simulations of the formation of star clusters, where the most massive stars typically form in the dense cluster cores (Klessen \etal 1998; Klessen and Burkert 2000, 2001; Bonnell and Bate 2002; Bonnell \etal 2003, 2007; Bate 2009a).  These features are in fact found in many simulations of cluster formation, including even the first crude ones by Larson (1978), and this suggests that they are a general result of the gravitational dynamics of collapsing and fragmenting clouds.  In the simulations of Bonnell and Bate (2002), Bonnell \etal (2003), and Bate (2009a), clusters form hierarchically by the merging of smaller stellar groupings in a process somewhat analogous to hierarchical galaxy formation.  At the same time, the gas and forming stars become increasingly concentrated in a dense core region where massive stars continue to gain mass by accretion.  All of these processes occur on a local dynamical timescale that decreases with increasing density, so that the formation of a dense cluster core and the growth of massive stars in it are both runaway processes.  If the stellar density becomes sufficiently high, for example in a tight binary or multiple system, stellar mergers may also occasionally occur and contribute to building the most massive stars (Bonnell \etal 1998; Bonnell and Bate 2005; Zinnecker and Yorke 2007).

   Models in which massive stars form by the collapse of separate massive cloud cores have also been studied using simulations that include radiative heating of the star-forming gas (Krumholz \etal 2007, 2009).  Since radiative heating tends to suppress fragmentation near a forming star, it might be expected to reduce the importance of interactions with nearby stars; however, close companions do still form in the simulations of Krumholz \etal (2007, 2009) and also in the simulation by Bate (2009b) including radiative heating, and these companions participate in strong dynamical interactions like those discussed above.  This occurs partly because, as was noted by Bate (2009b), gravitational drag effects can rapidly bring stars into tightly bound systems even if they initially formed at much larger separations.  The simulation of Krumholz \etal (2009) produces a massive and strongly interacting binary system along with several smaller companion objects, and this suggests that interactions with companions are likely to play an important role quite generally.

\subsection{Possible implications for the upper stellar IMF}  

   The observed fact that massive stars typically form near other massive stars suggests that their formation may be facilitated by interactions with other relatively massive stars, as might be expected if massive companions play a particularly important role in redistributing angular momentum during their formation.  Massive stars might then form by a bootstrap process that builds up stars of progressively larger mass (Larson 2002).  If this process is self-similar and yields a constant ratio in the numbers of stars in successive logarithmic mass intervals, the resulting upper Initial Mass Function will be a power law, consistent with observations (Salpeter 1955).  Unlike the earlier derivation of a power-law upper IMF by Zinnecker (1982) from Bondi-Hoyle accretion, a bootstrap model does not predict a runaway growth in mass of the most massive object or a growing gap in mass between the most massive and second most massive objects, but yields instead a continuous upper IMF because the numbers of stars in adjacent mass intervals are coupled.  The slope of the resulting power-law IMF cannot yet be predicted quantitatively, but numerical simulations of cluster formation incorporating the processes discussed here yield an upper IMF similar to the empirical Salpeter power law (Klessen 2001; Bonnell \etal 2003, 2007; Bate 2009a).

   Gravitational torques can be generated by collections of stars as well as by individual stars, and if clusters are built by the merging of subsystems, gravitational interactions among the subsystems may help to redistribute angular momentum and drive the gas flows that form massive stars, perhaps in a way analogous to the way that galaxy interactions  drive gas inflows in forming galaxies.  Because a larger system can redistribute more angular momentum, the mass of the most massive star that forms might be expected to increase systematically with the mass of the system, and this is indeed seen in the observations: with limited data, Larson (2003a) found that the mass of the most massive star in a young cluster increases with the mass of the cluster roughly as $M_{\rm cluster}^{0.45}$.  In a more complete study Weidner and Kroupa (2006) suggested that a better fit to the data is provided by a function with a logarithmic slope of 0.74 at small masses that flattens at high masses to an upper mass limit of $\sim 150 M_\odot$.  With additional data, Maschberger and Clarke (2008) found qualitatively similar trends, still with a large scatter, but they argued that no strong conclusion can yet be drawn about the form of the relation.
 
   If the mass of the most massive star in a cluster increases as a power of the cluster's mass, this also can generate a power-law upper IMF.  If all stars form in a self-similar clustering hierarchy and if the mass of the most massive star in each subcluster increases as a power $n$ of the subcluster's mass, then a power-law IMF $dN/d\log m \propto m^{-x}$ is produced with $x = 1/n$ (Larson 1991, 1992).  For example, a Salpeter upper IMF with $x = 1.35$ results if $n = 0.74$.  Such a hierarchical model for the origin of the upper IMF was proposed by Larson (1991) with no physical basis and elaborated by Larson (1992) on the basis of a fractal model, but we can now suggest a more physical basis for it involving the gravitational coupling effects discussed above.  As a result of these effects, a larger system can produce stronger torques and redistribute more angular momentum, allowing it to form more massive stars.

   Among well-studied clusters, the increase of maximum stellar mass with cluster mass continues up to a maximum stellar mass of $\sim 150 M_\odot$ in clusters with masses of order $10^5 M_\odot$ (Weidner and Kroupa 2006).  There is debated evidence that `intermediate-mass black holes' with masses of several thousand $M_\odot$ may exist in some of the most massive star clusters in the nearby universe with masses of order $10^6$ to $10^7 M_\odot$ (e.g., Maccarone and Servillat 2008), possibly continuing the trend of increasing maximum object mass with increasing cluster mass.  On still larger scales, there is no question that most galaxy bulges have supermassive central black holes whose masses are typically 0.001 to 0.002 times the bulge mass (Kormendy and Richstone 1995; Magorrian \etal 1998; Kormendy and Gebhardt 2001; H\"aring and Rix 2004; Ferrarese \etal 2006).  The ratio of black hole mass to bulge mass in galaxies is thus similar to the ratio of maximum stellar mass to cluster mass in the most massive star clusters.  This coincidence, while only in order of magnitude, suggests that qualitatively similar processes might operate in massive star clusters and galaxy bulges to build massive central objects, and this possibility will be discussed further in Section 5.

\subsection{Implications for the efficiency of star formation}  

   If star formation depends on the redistribution of angular momentum, the efficiency of star formation will depend on the effectiveness with which angular momentum can be redistributed in star-forming clouds.  If the redistribution is caused partly by gravitational interactions among the forming stars, as suggested above, then star formation should be most efficient in the densest environments where interactions are most important.  The simplest example would be a close binary system where the gas is converted efficiently into stars because of strong tidal effects that effectively redistribute angular momentum (Bate 2000).  The efficiency of star formation is also observed to be high in clusters, as would be expected if interactions among the forming stars play an important role; although the overall efficiency of star formation in molecular clouds is low, typically only a few percent (Myers \etal 1986; Evans and Lada 1991), the efficiency in forming clusters is about an order of magnitude higher than this, as would indeed be required if a newly formed cluster is to remain bound (Lada and Lada 2003).

\subsection{Summary}  

   Massive stars typically form near other massive stars in multiple systems and dense cluster cores, suggesting that the presence of massive companions in an associated cluster plays an important role in their formation.  The mass of the most massive star in a cluster might then be expected to increase systematically with the mass of the cluster, as is observed.  If clusters are built hierarchically and the mass of the most massive star increases as a power of the mass of the associated system, as observations suggest, a power-law upper IMF can be generated.  The formation of massive stars may then be closely linked to the formation of clusters, just as the formation of low-mass stars may be closely linked to the formation of binary or multiple systems.  Star formation may thus generally be a collective enterprise.

\section{The formation of supermassive black holes in galaxies}  

   Much evidence shows that nearly all galaxies with a significant bulge host a supermassive central black hole.  As is true for star formation, the biggest obstacle to the formation of these central black holes is the requirement that the matter from which they form must lose nearly all of its initial angular momentum during the formation process (Shlosman \etal 1990; Phinney 1994; Escala 2006).  In Sections 3 and 4 it was argued that the formation of stars is tied to the formation of an associated stellar system that plays an important role in redistributing the excess angular momentum involved.  The formation of black holes in galactic nuclei may similarly be coupled to the formation of the galaxies that host them, as is suggested by the close correlation between black hole mass and bulge mass (Heckman \etal 2004).  The observed relation between black hole mass and bulge mass (Kormendy and Richstone 1995; Magorrian \etal 1998; Kormendy and Gebhardt 2001; H\"aring and Rix 2004; Ferrarese \etal 2006) resembles the relation between maximum stellar mass and cluster mass discussed in Section 4.2, and the two relations intersect at an intermediate mass when extrapolated, suggesting the possibility that there might be a continuity between them and that similar processes might be involved in establishing them (Larson 2003a,b).

   Central black holes in galaxies can gain mass in two ways: smaller black holes can merge into larger ones when galaxies merge, and an existing black hole can continue to grow by gas accretion.  Accretion may dominate the growth of moderate-mass black holes at high redshift, while merging may dominate the growth of massive black holes at low redshift (Malbon \etal 2007).  In both cases angular momentum must be removed, either from the orbits of the merging black holes or from the gas being accreted by a central black hole.  As is the case with star formation, the growth of a central black hole in a galaxy may partly involve disk processes, but it is also likely to involve gravitational interactions with massive objects or mass concentrations the surrounding system.

\subsection{The role of disks}  

   Accretion disks are thought to play an important role in feeding the central black holes in galaxies, and evidence for this is provided by the fact that AGNs often have jets believed to be powered by an accretion disk around the central black hole.  Standard models postulate a compact thin accretion disk with a radius of order 0.1 parsecs in which some source of effective viscosity such as MHD turbulence drives an inflow (Frank \etal 2002; King \etal 2008).  Such compact disks have not been directly observed, but in a few cases radio maser observations have revealed the existence of somewhat larger disks with radii of about 1 parsec around the central black holes of some galaxies (Kondratko \etal 2008).  In the best studied case of NGC 4258, this disk is perpendicular to a radio jet, as would be expected if the jet is powered by accretion from the disk.

   At radii much larger than 0.1 parsecs, standard accretion disk models become gravitationally unstable, and their relevance then becomes questionable because they may form stars instead of feeding a central black hole (Shlosman and Begelman 1989; Begelman 1994; Goodman 2003; Tan and Blackman 2005; King \etal 2008).  The disk inflow time also becomes very long at these radii, longer than 1 Gyr and therefore too long to be compatible with the timescale of AGN activity.  The question then becomes how gas can be fed efficiently from regions of galactic size into a compact accretion disk only 0.1 parsecs in radius around the black hole (Phinney 1994; Escala 2006).  One possible source of gas to feed a central black hole and its compact accretion disk might be a much larger and more massive disk several hundred parsecs in radius, like the nuclear gas disks seen in ULIRG/starburst galaxies (Downes and Solomon 1998; Riechers \etal 2009) and in simulations of galaxy mergers (Barnes and Hernquist 1996; Barnes 2002; Mayer \etal 2009).  These disks are massive enough to be gravitationally unstable, and the gravitational torques created by the resulting density fluctuations can drive some of the gas toward the center (Shlosman and Begelman 1989; Shlosman 1992; Escala 2006, 2007; Levine \etal 2008).  Such disks may also play a role in the merging of black holes in galactic nuclei (Escala \etal 2005; Dotti \etal 2006, 2007; Cuadra \etal 2009), in which  case the orbiting black holes may accrete gas from the disk while spiraling inward (Escala 2004; Dotti \etal 2007).

   Simulations of massive nuclear gas disks several hundred parsecs in size show that they fragment rapidly into massive clumps and filaments that may both form stars and drive an inflow toward the center (Wada and Norman 2002; Wada 2004; Escala 2007; Escala and Larson 2008).  Star formation and AGN activity may then be closely linked because they have the same basic cause, namely gravitational instability in a massive nuclear gas disk.  Star formation in such disks may be concentrated in very massive clusters (Escala and Larson 2008), and these massive clusters might also contribute to the redistribution of angular momentum in the system, perhaps playing a role analogous to that of companion stars in star formation, as discussed below.

\subsection{The role of mass concentrations and asymmetries in the surrounding system}  

   A clue to the nature of the black hole accretion process is provided by the fact that the spin axes of the central black holes in galaxies as inferred from AGN jet directions are randomly oriented and not correlated with the large-scale structure of the galaxy (Kinney \etal 2000).  This suggests that central black holes are not fed by a smooth inflow from a disk but by a more irregular processes such as the infall of dense clumps of gas (Nayakshin and King 2009).  The associated redistribution of angular momentum must then also be a rather irregular process, involving for example gravitational interactions between the infalling clumps and other massive objects or mass concentrations in the system.

   Massive objects or mass concentrations that could help to redistribute angular momentum are indeed observed in galactic nuclei.  The central part of our Galaxy contains a number of massive GMCs and star clusters, including the Arches and Quintuplet clusters within a few tens of parsecs of the central black hole, and most notably the Galactic Center Cluster surrounding it within 1 parsec (Morris and Serabyn 1995).  The Galactic Center Cluster may itself consist of three subsystems, two nearly orthogonal disks or rings and a third more compact system within 0.1 parsec of the black hole (Paumard \etal 2006; Genzel and Karas 2007; Sch\"odel \etal 2007; L\"ockmann and Baumgardt 2009).  All of these systems are only several Myr old, implying that they must have formed not far from where they are.  The stars in these systems could have formed from accretion disks around the black hole (Levin and Beloborodov 2003; Nayakshin \etal 2007; Alexander \etal 2008), or they could have formed from gas clouds captured into orbit around it (Wardle and Yusef-Zadeh 2008; Bonnell and Rice 2008; Mapelli \etal 2009).  In either case, as noted by these authors, some of the gas from which these systems formed would almost certainly have gone into the central black hole.  Some of the angular momentum of the gas that feeds the central black hole might then end up in the orbital motions of the stars around it, as happens with star formation.

   Just outside the Galactic Center Cluster there is a ring of dense clumps of molecular gas with a total mass of about $10^6\,M_\odot$ and a peak density located at about 1.6 parsecs from the center (Christopher et al.\ 2005).  These clumps have individual masses of a few times $10^4\,M_\odot$ and are dense enough to be gravitationally bound against tidal forces, so they could plausibly provide a source of gas for star formation in the Galactic Center Cluster (Christopher \etal 2005).  Some of the gas in this ring may fall inward and fuel star formation or feed the central black hole if this gas loses angular momentum by gravitational interaction with the dense clumps or with other mass concentrations in the region.

   The Galactic Center Cluster is not unique to our Galaxy, since a very compact cluster of young stars is also seen around the central black hole of M31, where it forms part of a larger stellar nucleus that is flattened and rotating (Bender \etal 2005).  Distinct stellar nuclei are also seen in many other galaxies; in fact, nearly all galaxies of small and intermediate mass have central light concentrations or `nuclei' several parsecs to several tens of parsecs in radius (Rossa \etal 2006; C\^ot\'e \etal 2006, 2007; Ferrarese \etal 2006).  These stellar nuclei, also called nuclear star clusters, are similar in size to globular clusters but many times more luminous.  Remarkably, they follow the same relation between central object mass and bulge mass as do the central black holes in galaxies: for both stellar nuclei and black holes, the mass of the central object is typically about 0.001 to 0.002 times the mass of the  bulge (Wehner and Harris 2006; Ferrarese \etal 2006).  Some galaxies have both a stellar nucleus and a black hole, and in these cases the two are of comparable mass (Seth \etal 2008); in M31, the stellar nucleus has a mass of a few times $10^7\,M_\odot$, a few times smaller than the black hole mass of about $10^8\,M_\odot$ (Peiris and Tremaine 2003; Bender \etal 2005).  Although stellar nuclei are not seen in the most massive galaxies, they could have existed in the smaller progenitor galaxies that merged to form them and could have been destroyed by the mergers.

   The apparent close correspondence between stellar nuclei and central black holes in galaxies suggests that their formation processes may be linked.  One possibility is that a compact nucleus forms first by the accumulation of matter at the center of a galaxy, and that efficient redistribution of angular momentum in this nucleus then allows a significant fraction of its mass to go into a central black hole.  In simulations of cluster formation by gravitational instability in galactic gas disks, Li \etal (2007) found that the most massive cluster forms at the center or soon settles there by dynamical friction, and that the mass of this central cluster increases with galaxy mass in a way that resembles the observed relation between nuclear mass and galaxy mass (Wehner and Harris 2006; Ferrarese \etal 2006).  A similar tendency for massive clumps formed by gravitational instability in the disk of a young galaxy to sink rapidly toward the center was also found in the simulations of Elmegreen \etal (2008a,b).

   A more organized gravitational phenomenon that can drive gas inflows in galaxies is the occurrence of collective effects such disk asymmetries and bars.  The nucleus of M31 has an asymmetric and double-peaked light profile in which the more prominent peak P1 is offset by about 2 pc from the less prominent one P2, which is centered on the black hole and contains a more compact young cluster at its center (Bender \etal 2005).  The off-center peak P1 can be explained by an eccentric disk of stars orbiting around the black hole in such a way that the stars linger near the outer ends of their orbits and create a mass concentration there, about 2 pc from the black hole (Peiris and Tremaine 2003).  The resulting asymmetric mass distribution may then exert a torque on any gas in the nucleus and drive an inflow toward the black hole (Chang \etal 2007).  On larger scales, galactic bars can be very effective in extracting angular momentum from the gas in galaxies and driving it toward the center (Athanassoula 1992, 1994; Kenney 1994; Kennicutt 1994; Sellwood and Shen 2004).  Such bars can drive gas into the inner several hundred parsecs of a galaxy, and smaller `bars within bars' have been suggested to drive inflows on smaller scales (Shlosman \etal 1989, 1990; Englmaier and Shlosman 2004; Heller \etal 2007; Namekata \etal 2009), although this possibility has been disputed (Maciejewski \etal 2002).  Another possibility is that nuclear spirals may drive inflows that feed the central black holes in galaxies (van de Ven and Fathi 2009).  In addition to large-scale bars, asymmetric or lopsided disks are also common in galaxies and can also drive inflows (Jog and Combes 2009; Reichard \etal 2009).

\subsection{Star formation and black hole feeding}  

   Observations suggest a connection between AGN activity and recent star formation activity in galaxies (Kauffmann \etal 2003; Davies \etal 2007; Watabe \etal 2008; Reichard \etal 2009), and there is also a good correspondence between the cosmic history of AGN activity and the cosmic history of star formation (Somerville \etal 2008).  A connection might be expected because the same gas that fuels an episode of star formation in a galactic nucleus may also feed a central black hole.  Star formation very close to a massive black hole has been considered problematic because the strong tidal forces present would quickly disrupt star-forming clouds like those observed in the Solar vicinity.  To survive tidal disruption and form stars, a cloud near a massive black hole would have to have an exceptionally high density, but the clouds in the Galactic Center region do in fact have exceptionally high densities, several orders of magnitude higher than those of nearby molecular clouds (Morris and Serabyn 1996; Christopher and Scoville 2003; G\"usten and Philipp 2004; Christopher \etal 2005).  Simulations of infalling gas clouds or steams around a central black hole show that some of the infalling gas forms disks or rings around the black hole in which the density can become high enough to overcome tidal effects and form stars (Bonnell and Rice 2008; Mapelli \etal 2009).  If star formation does occur in dense gas orbiting around a central black hole, the same gravitational instabilities that form the stars may also help to drive an inflow toward the black hole.

   Simulations of more massive and extended nuclear gas disks by Escala (2007) show that such disks can develop large density fluctuations that simultaneously form stars and drive an inflow toward a central black hole.  These simulations also show the formation of massive clumps with masses of several times $10^6\,M_\odot$, about the mass predicted by stability theory for the largest self-gravitating structures not stabilized by rotation (Escala and Larson 2008).  With a plausible formation efficiency, such clumps might form star clusters with the masses of globular clusters, and such clusters are indeed seen in starburst systems such as NGC 4038/39 (Escala and Larson 2008).  As was noted by these authors, even more massive clumps are predicted to form in more gas-rich systems, and these very massive clumps might contribute importantly to the redistribution of angular momentum; it is also possible that some of them might lose enough angular momentum by gravitational drag to sink toward the center like the `giant clumps' in the simulations of Elmegreen et al (2008a,b), where they might help to build up a stellar nucleus and central black hole.  Again, this would imply a close connection between star formation and black hole feeding in galaxies.

\subsection{Possible implications for black hole masses}  

   It was suggested in Sections 3 and 4 that the gravitational interactions that redistribute angular momentum in a system of forming stars may couple the mass of the most massive forming object to the mass of the system, and it was noted that this can have important implications for the mass ratios of binaries and the masses of the most massive stars in clusters.  Some of the gravitational effects that redistribute angular momentum in a galaxy with a growing central black hole might similarly have a tendency to couple the mass of the black hole to the mass of the galaxy or its nucleus.

   If asymmetric nuclear disks like that in M31 are common and play a role in feeding central black holes (Chang \etal 2007), the mass of the black hole and the mass of the disk around it may be coupled.  Although the nucleus of M31 is dominated by the black hole and the orbits of the stars are therefore almost Keplerian, the eccentric disk must still have enough self-gravity to keep the orbits aligned, and this implies that it cannot be too much less massive than the black hole.  This requirement may regulate the growth of the black hole and prevent it from becoming much more massive than the nuclear disk.  Such effects could produce a tendency for the mass of the central black hole in a galaxy to equalize with the mass of the surrounding stellar nucleus, perhaps in a way analogous to the tendency for the masses of the stars in close binaries to equalize.

  Similar self-regulating effects might also, at least in principle, act on larger scales to limit bar-driven inflows, because if a central object becomes too massive it can weaken or destroy a bar and thus shut off the inflow.  According to Hozumi and Hernquist (2005), a central object with a mass only 0.5\% of the disk mass can destroy a bar in a short time, and an object with a mass as small as 0.2\% of the disk mass can significantly weaken a bar over a Hubble time.  Hozumi and Hernquist note that these masses are within the range of observed black hole masses in galaxies, and they suggest that bar destruction by central black holes might therefore indeed occur in some galaxies.  However, this conclusion is very uncertain because other studies have found that a much larger central mass, perhaps an order of magnitude larger, is needed to destroy a bar (Shen and Sellwood 2004; Athanassoula et al 2005), and in this case bar destruction is unlikely to occur by this mechanism.  Therefore it is not presently clear whether gravitational effects alone can establish a relation between black hole mass and galaxy mass, although these effects can nevertheless play an important role in establishing the radial mass distribution in galaxies, and this radial mass distribution must in turn be relevant to the problem of understanding black hole masses.

   Explanations for the relation between black hole mass and bulge mass based on AGN feedback effects have also been proposed, and have received much attention in the literature (e.g., Robertson \etal 2006; Di Matteo \etal 2008, Somerville \etal 2008; Younger \etal 2008).  These models postulate that energy from an accreting black hole can eventually expel the remaining gas from a galaxy and thereby terminate the black hole's growth.  However, the physics involved is complex and poorly understood and represented by adjustable parameters.  Thus without a much more quantitative understanding of both feedback effects and angular momentum transport, it will not be possible to decide which is more important in regulating the growth of black holes.

\subsection{Summary}  

   As is the case with star formation, most of the initial angular momentum of the matter from which a central black hole forms must be removed during the formation process, and gravitational interactions with mass concentrations in the surrounding galaxy can play an important role.  Several effects, including gravitational instability in a nuclear gas disk, interaction with nearby massive objects or mass concentrations, and collective phenomena such a galactic bars or disk asymmetries, may all contribute.  Much of the angular momentum of the matter that goes into the central black hole might then end up in the orbital motions of stars in the galaxy.  As with star formation, these processes may tend to couple the mass of the central object to the mass of the surrounding system, but a full understanding of this problem almost certainly involves other physical effects as well and awaits further work.

\section{Conclusions}  

   The biggest obstacle to the formation of compact objects like stars and black holes from diffuse matter is that because of their tiny size in relation to the size of the system in which they form, they can absorb only a tiny fraction of the initial angular momentum of the matter from which they form.  Nearly all of this angular momentum must therefore be removed or redistributed during the formation process.  Magnetic braking can remove angular momentum from a diffuse star-forming cloud, and a rotating magnetically driven wind or jet can remove angular momentum from a newly formed compact object, but this leaves a large intermediate range of densities where the gas must lose at least 99\% of its angular momentum by non-magnetic processes.  In this paper it has been argued that the gravitational torques produced by companion stars and other mass concentrations in a forming system may account for much of the redistribution of angular momentum in this regime.

   For low-mass stars that typically form in binary or multiple systems, the companion stars may take up most of the angular momentum of the gas that goes into each star and transfer some of it to outlying gas.  For massive stars that typically form in compact multiple systems and clusters, interactions with neighboring massive stars or stellar groupings may play a similar role.  For supermassive black holes that form at the centers of galaxies, several gravitational effects may contribute, including gravitational instability in a gas disk, interactions with mass concentrations in the surrounding system, and the effect of bars and disk asymmetries in driving gas inflows.  These effects may generally tend to couple the mass of the forming object to that of the system in which it forms.

   If gravitational interactions with companion objects or other mass concentrations in an associated system play an important role in the formation of both stars and black holes, these objects may rarely form in isolation and their formation may generally be a collective enterprise involving interactions with a larger system.  The formation of stars and black holes will then often be a more complex, dynamic, and chaotic process than in standard models, and detailed numerical simulations are needed to gain further insight into these processes (Larson 2007).

\References

\item[] Abt H A 1983 {\it Ann.\ Rev.\ Astron.\ Astrophys.} {\bf 21}
  343--72

\item[] Abt H A, Gomez A E, and Levy S G 1990 {\it Astrophys.\ J.\
  Suppl.} {\bf 74} 551--73

\item[] Alexander R D, Armitage P J, Cuadra J, and Begelman M C 2008 {\it
  Astrophys.\ J.} {\bf 674} 927--35

\item[] Arce H G, Shepherd D, Gueth F, Lee C-F, Bachiller R, Rosen A and
  Beuther H 2007 {\it Protostars and Planets V} ed B Reipurth {\it et
  al} (Tucson: University of Arizona Press) pp 245--60

\item[] Artymowicz P 1983 {\it Acta Astronomica} {\bf 33} 223--41

\item[] Athanassoula E 1992 {\it Mon.\ Not.\ R. Astron.\ Soc.} {\bf 259}
  345--64

\item[] Athanassoula E 1994 {\it Mass-Transfer Induced Activity in
  Galaxies} ed I Shlosman (Cambridge: Cambridge University Press)
  pp 143--54

\item[] Athanassoula E, Lambert J C and Dehnen W 2005 {\it Mon.\ Not.\
  R. Astron.\ Soc.} {\bf 363} 496--508

\item[] Attwood R E, Goodwin S P, Stamatellos D and Whitworth A P 2009
  {\it Astron.\ Astrophys.} {\bf 495} 201--15

\item[] Banerjee R and Pudritz R E 2006 {\it Astrophys.\ J.} {\bf 641}
  949--60

\item[] Barnes J E 2002 {\it Mon.\ Not.\ R. Astron.\ Soc.} {\bf 333}
  481--94

\item[] Barnes J E and Hernquist L 1996 {\it Astrophys.\ J.} {\bf 471}
  115--42

\item[] Barranco J A and Goodman A A 1998 {\it Astrophys.\ J.} {\bf 504}
  207--22

\item[] Basu S 1997 {\it Astrophys.\ J.} {\bf 485} 240--53

\item[] Basu S and Mouschovias T Ch 1994 {\it Astrophys.\ J.} {\bf 432}
  720--41

\item[] Basu S and Mouschovias T Ch 1995 {\it Astrophys.\ J.} {\bf 452}
  386--400

\item[] Bate M R 1997 {\it Mon.\ Not.\ R. Astron.\ Soc.} {\bf 285} 16--32

\item[] Bate M R 2000 {\it Mon.\ Not.\ R. Astron.\ Soc.} {\bf 314} 33--53

\item[] Bate M R 2009a {\it Mon.\ Not.\ R. Astron.\ Soc.} {\bf 392}
  590--616

\item[] Bate M R 2009b {\it Mon.\ Not.\ R. Astron.\ Soc.} {\bf 392}
  1363--80

\item[] Bate M R and Bonnell I A 1997 {\it Mon.\ Not.\ R. Astron.\ Soc.}
  {\bf 285} 33--48

\item[] Bate M R and Bonnell I A 2005 {\it Mon.\ Not.\ R. Astron.\ Soc.}
  {\bf 356} 1201--21

\item[] Bate M R, Bonnell I A and Bromm V 2002 {\it Mon.\ Not.\ R.
  Astron.\ Soc.} {\bf 332} L65--8

\item[] Bate M R, Bonnell I A and Bromm V 2003 {\it Mon.\ Not.\ R.
  Astron.\ Soc.} {\bf 339} 577--99

\item[] Begelman M C 1994 {\it Mass-Transfer Induced Activity in
  Galaxies} ed I Shlosman (Cambridge: Cambridge University Press)
  pp 23--9

\item[] Belloche A, Andr\'e P, Despois D and Blinder S 2002 {\it
  Astron.\ Astrophys.} {\bf 393} 927--47

\item[] Bender R, Kormendy J, Bower G, Green R, Thomas J, Danks A C,
  Gull T, Hutchings J B, Joseph C L, Kaiser M E, Lauer T, Nelson C H,
  Richstone D, Weistrop D and Woodgate B 2005 {\it Astrophys.\ J.}
  {\bf 631} 280--300 

\item[] Beuther H, Churchwell E B, McKee C F and Tan J C 2007 {\it
  Protostars and Planets V} ed B Reipurth \etal (Tucson: University
  of Arizona Press) pp 165--80  

\item[] Blondin J M 2000 {\it New Astron.} {\bf 5} 53--68 

\item[] Bodenheimer P 1995 {\it Annu.\ Rev.\ Astron.\ Astrophys.}
  {\bf 33} 199--238

\item[] Bodenheimer P, Ruzmaikina T and Mathieu R D 1993 {\it Protostars
  and Planets III} ed E H Levy and J I Lunine (Tucson: University of
  Arizona Press) pp 367--404

\item[] Bodenheimer P, Burkert A, Klein R I and Boss A P 2000 {\it
  Protostars and Planets IV} ed V Mannings \etal (Tucson: University of
  Arizona Press) pp 675--701

\item[] Boffin H M J 2001 {\it Astrotomography: Indirect Imaging Methods
  in Observational Astronomy (Lecture Notes in Physics Vol.\ 573)}
  ed H M J Boffin \etal (Berlin: Springer) p 69--87

\item[] Bonnell I and Bastien P 1992 {\it Astrophys.\ J.} {\bf 401}
  L31--4

\item[] Bonnell I A and Bate M R 1994 {\it Mon.\ Not.\ R. Astron.\ Soc.}
  {\bf 271} 999--1004

\item[] Bonnell I A and Bate M R 2002 {\it Mon.\ Not.\ R. Astron.\ Soc.}
  {\bf 336} 659--69

\item[] Bonnell I A and Bate M R 2005 {\it Mon.\ Not.\ R. Astron.\ Soc.}
  {\bf 362} 915--20

\item[] Bonnell I A and Rice W K M 2008 {\it Science} {\bf 231} 1060--82

\item[] Bonnell I A, Bate M R and Zinnecker H 1998 {\it Mon.\ Not.\ R.
  Astron.\ Soc.} {\bf 298} 93--102

\item[] Bonnell I A, Bate M R and Vine S G 2003 {\it Mon.\ Not.\ R.
  Astron.\ Soc.} {\bf 343} 413--8

\item[] Bonnell I A, Larson R B and Zinnecker H 2007 {\it Protostars and
  Planets V} ed B. Reipurth \etal (Tucson: University of Arizona Press)
  pp 149--64

\item[] Boss A P 2002 {\it Astrophys.\ J.} {\bf 576} 462--72

\item[] Caselli P, Benson P J, Myers P C and Tafalla M 2002 {\it
  Astrophys.\ J.} {\bf 572} 238--63

\item[] Cesaroni R, Galli D, Lodato G, Walmsley C M and Zhang Q 2007
  {\it Protostars and Planets V} ed B Reipurth \etal (Tucson:
  University of Arizona Press) pp 197--213 

\item[] Chang P, Murray-Clay R, Chiang E and Quataert E 2007 {\it
  Astrophys.\ J.} {\bf 668} 236--44

\item[] Christopher M H and Scoville N Z 2003 {\it Active Galactic
  Nuclei: From Central Engine to Galaxy (ASP Conf.\ Ser.\ Vol.\ 290)}
  ed S Collin \etal (San Francisco: Astronomical Society of the Pacific)
  pp 389--90

\item[] Christopher M H, Scoville N Z, Stolovy S R and Yun M S 2005 {\it
  Astrophys.\ J.} {\bf 622} 346--65 

\item[] Coffey D, Bacciotti F, Ray T P, Eisl\"offel J and Woitas J 2007
  {\it Astrophys.\ J.} {\bf 663} 350--64 

\item[] Coffey D, Bacciotti F and Podio L 2008 {\it Astrophys.\ J.}
  {\bf 689} 1112--26

\item[] C\^ot\'e P, Piatek S, Ferrarese L, Jord\'an A, Merritt D, Peng
  E W, Ha\c segan M, Blakeslee J P, Mei S, West M J, Milosavlevi\'c M
  and Tonry J L 2006 {\it Astrophys.\ J.\ Suppl.} {\bf 165} 57--94 

\item[] C\^ot\'e P, Ferrarese L, Jord\'an A, Blakeslee J P, Chen C-W,
  Infante L, Merritt D, Mei S, Peng E W, Tonry J L, West A W and West M J
  2007 {\it Astrophys.\ J.} {\bf 671} 1456--65

\item[] Crutcher R M 1999 {\it Astrophys.\ J.} {\bf 520} 706--13

\item[] Cuadra J, Armitage P J, Alexander R D, Begelman M C 2009 {\it
  Mon.\ Not.\ R. Astron.\ Soc.} {\bf 393} 1423--32

\item[] Davies R I, Mueller S\'anchez F, Genzel R, Tacconi L J, Hicks
  E K S and Friedrich S 2007 {\it Astrophys.\ J.} {\bf 671} 1388--1412

\item[] Delgado-Donate E J, Clarke C J, Bate M R and Hodgkin S T 2004
  {\it Mon.\ Not.\ R. Astron.\ Soc.} {\bf 351} 617--29

\item[] de Wit W J, Testi L, Palla F and Zinnecker H 2005 {\it Astron.\
  Astrophys.} {\bf 437} 247--55

\item[] Di Matteo T, Colberg J, Springel V, Hernquist L and Sijacki D
  2008 {\it Astrophys.\ J.} {\bf 676} 33--53

\item[] Dotti M, Colpi M and Haardt F 2006 {\it Mon.\ Not.\ R. Astron.\
  Soc.} {\bf 367} 103--12

\item[] Dotti M, Colpi M, Haardt F and Mayer L 2007 {\it Mon.\ Not.\ R.
  Astron.\ Soc.} {\bf 379} 956--62

\item[] Downes D and Solomon P M 1998 {\it Astrophys.\ J.} {\bf 507}
  615--54

\item[] Duch\^ene G, Bontemps S, Bouvier J, Andr\'e P, Djupvik A A and
  Ghez A M 2007 {\it Astron.\ Astrophys.} {\bf 476} 229--42

\item[] Duquennoy A and Mayor M 1991 {\it Astron.\ Astrophys.} {\bf 248}
  485--524

\item[] Elmegreen B G, Bournaud F, and Elmegreen D M 2008a {\it
  Astrophys.\ J.} {\bf 684} 829--34

\item[] Elmegreen B G, Bournaud F, and Elmegreen D M 2008b {\it
  Astrophys.\ J.} {\bf 688} 67--77

\item[] Englmaier P and Shlosman I 2004 {\it Astrophys.\ J.} {\bf 617}
  L115--8

\item[] Escala A 2004 {\it Ph.D. Thesis, Yale University}

\item[] Escala A 2006 {\it Astrophys.\ J.} {\bf 648} L13--6

\item[] Escala A 2007 {\it Astrophys.\ J.} {\bf 671} 1264--71

\item[] Escala A and Larson R B 2008 {\it Astrophys.\ J.} {\bf 685}
  L31--4

\item[] Escala A, Larson R B, Coppi P S, and Mardones D 2005 {\it
  Astrophys.\ J.} {\bf 630} 152--66

\item[] Evans N J and Lada E A 1991 {\it Fragmentation of Molecular
  Clouds and Star Formation (IAU Symp.\ No.\ 147)} ed E Falgarone
  \etal (Dordrecht: Kluwer) pp 293--315

\item[] Ferrarese L, C\^ot\'e P, Bont\`a E D, Peng E W, Merritt D,
  Jord\'an A, Blakeslee J P, Ha\c segan M, Mei S, Piatek S, Tonry J L
  and West M J 2006 {\it Astrophys.\ J.} {\bf 644} L21--4

\item[] Fischer D A and Marcy G W 1992 {\it Astrophys.\ J.} {\bf 396}
  178--84

\item[] Fisher R T 2004 {\it Astrophys.\ J.} {\bf 600} 769--80

\item[] Frank J, King A and Raine D J 2002 {\it Accretion Power in
  Astrophysics} (Cambridge: Cambridge University Press)

\item[] Genzel R and Karas V 2007 {\it Black Holes from Stars to Galaxies
  -- Across the Range of Masses (IAU Symp.\ No.\ 238)} ed V Karas
  \etal (Cambridge: Cambridge University Press) pp 173--80 

\item[] Goldberg D, Mazeh T and Latham D W 2003 {\it Astrophys.\ J.}
  {\bf 591} 397--405

\item[] Goodman J 2003 {\it Mon.\ Not.\ R. Astron.\ Soc.} {\bf 339}
  937--48

\item[] Goodman A A, Benson P J, Fuller G A and Myers P C 1993 {\it
  Astrophys.\ J.} {\bf 406} 528--47

\item[] Goodwin S P and Kroupa P 2005 {\it Astron.\ Astrophys.} {\bf 439}
  565--9

\item[] Goodwin S P, Kroupa P, Goodman A and Burkert A 2007 {\it Protostars
  and Planets V} ed B Reipurth \etal (Tucson: University of
  Arizona Press) pp 133--28

\item[] G\"usten R and Philipp S D 2004 {\it The Dense Interstellar
  Medium in Galaxies} ed S Pfalzner \etal (Berlin: Springer) p 253

\item[] H\"aring N and Rix H-W 2004 {\it Astrophys.\ J.} {\it 604}
  L89--92

\item[] Hartmann L 1998 {\it Accretion Processes in Star Formation}
  (Cambridge: Cambridge University Press)

\item[] Heckman T M, Kauffmann G, Brinchmann J, Charlott S, Tremonti C
  and White S D M 2004 {\it Astrophys.\ J.} {\bf 613} 109--18

\item[] Heiles C, Goodman ANA, McKee C F and Zweibel E G {\it Protostars
  and Planets III} ed E H Levy and J I Lunine (Tucson: University of
  Arizona Press) pp 279--326

\item[] Heintz W D 1978 {\it Double Stars} (Dordrecht: Reidel)

\item[] Heller C H 1993 {\it Astrophys.\ J.} {\bf 408} 337--46

\item[] Heller C H 1995 {\it Astrophys.\ J.} {\bf 455} 252--9

\item[] Heller C H, Shlosman I and Athanassoula E 2007 {\it Astrophys.\
  J.} {\bf 657} L65--8

\item[] Hennebelle P and Teyssier R 2008 {\it Astron.\ Astrophys.}
  {\bf 477} 25--34 

\item[] Hozumi S and Hernquist L 2005 {\it Publ.\ Astron.\ Soc.\ Japan}
  {\bf 57} 719--31

\item[] Jappsen A-K and Klessen R S 2004 {\it Astron.\ Astrophys.}
  {\bf 423} 1--12

\item[] Jijina J, Myers P C and Adams F C 1999 {\it Astrophys.\ J.\
   Suppl.} {\bf 125} 161--236

\item[] Jog C J and Combes F 2009 {\it Physics Reports} {\bf 471} 75--111

\item[] Kauffmann G, Heckman T M, Tremonti C, Brinchmann J, Charlot S,
  White S D M, Ridgway S E, Brinkmann J, Fukugita M, Hall P B, Ivezi\'c
  Z, Richards G T and Schneider D P 2003 {\it Mon.\ Not.\ R. Astron.\
  Soc.} {\bf 346} 1055--77

\item[] Kenney J D P 1994 {\it Mass-Transfer Induced Activity in
  Galaxies} ed I Shlosman (Cambridge: Cambridge University Press) pp 78--89 

\item[] Kennicutt R 1994 {\it Mass-Transfer Induced Activity in Galaxies}
  ed I Shlosman (Cambridge: Cambridge University Press) pp 131--42

\item[] Kinney A L, Schmitt H R, Clarke C J, Pringle J E, Ulvestad J S
  and Antonucci R R J 2000 {\it Astrophys.\ J.} {\bf 537} 152--77

\item[] King A R, Pringle J E and Hofmann J A 2008 {\it Mon.\ Not.\ R.
  Astron.\ Soc.} {\bf 385} 1621--7

\item[] Klessen R S 2001 {\it Astrophys.\ J.} {\bf 556} 837--46

\item[] Klessen R S and Burkert A 2000 {\it Astrophys.\ J. Suppl.}
  {\bf 128} 287--319

\item[] Klessen R S and Burkert A 2001 {\it Astrophys.\ J.} {\bf 549}
  386--401

\item[] Klessen R S, Burkert A and Bate M R 1998 {\it Astrophys.\ J.}
  {\bf 501} L205--8

\item[] Kobulnicky H A and Fryer C L 2007 {\it Astrophys.\ J.} {\bf 670}
  747--65

\item[] Kondratko P T, Greenhill L J and Moran J M 2008 {\it Astrophys.\
  J.} {\bf 678} 87--95

\item[] Kormendy J and Richstone D 1995 {\it Annu.\ Rev.\ Astron.\
  Astrophys.} {\bf 33} 581--624

\item[] Kormendy J and Gebhardt K 2001 {\it 20th Texas Symposium on
  Relativistic Astrophysics (AIP Conf.\ Proc.\ Vol.\ 586)} ed J C Wheeler
  \etal (New York: Springer) pp 363--81

\item[] Krumholz M R and Bonnell I A 2009 {\it Structure Formation in
  Astrophysics} ed G Chabrier (Cambridge: Cambridge University Press)
  pp 288--320

\item[] Krumholz M R, Klein R I and McKee C F 2007 {\it Astrophys.\ J.}
  {\bf 656} 959--79

\item[] Krumholz M R, Klein R I, McKee C F, Offner S S R and Cunningham
  A J 2009 {\it Science} {\bf 323} 754--7

\item[] Lada C J 2006 {\it Astrophys.\ J.} {\bf 640} L63--6 

\item[] Lada C J and Lada E A 2003 {\it Annu.\ Rev.\ Astron.\
  Astrophys.} {\bf 41} 57--115

\item[] Larson R B 1972 {\it Mon.\ Not.\ R. Astron.\ Soc.} {\bf 156}
  437--58

\item[] Larson R B 1978 {\it Mon.\ Not.\ R. Astron.\ Soc.} {\bf 184}
  69--85

\item[] Larson R B 1984 {\it Mon.\ Not.\ R. Astron.\ Soc.} {\bf 206}
  197--207

\item[] Larson R B 1989 {\it The Formation and Evolution of Planetary
  Systems} ed H A Weaver and L Danly (Cambridge: Cambridge University
  Press) pp 31--54

\item[] Larson R B 1990a {\it Physical Processes in Fragmentation and
  Star Formation} ed R Capuzzo-Dolcetta \etal (Dordrecht: Kluwer)
  pp 389--400

\item[] Larson R B 1990b {\it Mon.\ Not.\ R. Astron.\ Soc.} {\bf 243}
  588--92

\item[] Larson R B 1991 {\it Fragmentation of Molecular Clouds and Star
  Formation (IAU Symp.\ No.\ 147)} ed E Falgarone \etal (Dordrecht:
  Kluwer) pp 261--73

\item[] Larson R B 1992 {\it Mon.\ Not.\ R. Astron.\ Soc.} {\bf 256}
  641--6

\item[] Larson R B 2001 {\it The Formation of Binary Stars (IAU Symp.\
  No.\ 200)} ed H Zinnecker and R D Mathieu (San Francisco: Astronomical
  Society of the Pacific) pp 93--106

\item[] Larson R B 2002 {\it Mon.\ Not.\ R. Astron.\ Soc.} {\bf 332}
  155--64

\item[] Larson R B 2003a {\it Galactic Star Formation Across the Stellar
  Mass Spectrum (ASP Conf.\ Ser.\ Vol.\ 287)} ed J M De Buizer and
  N S van der Bliek (San Francisco: Astronomical Society of the Pacific)
  pp 65--80

\item[] Larson R B 2003b {\it Rep.\ Prog.\ Phys.} {\bf 66} 1651--97

\item[] Larson R B 2007 {\it Rep.\ Prog.\ Phys.} {\bf 70} 337--56

\item[] Levin Y and Beloborodov A M 2003 {\it Astrophys.\ J.} {\bf 590}
  L33--6

\item[] Levine R, Gnedin N Y, Hamilton A J S and Kravtsov A V 2008 {\it
  Astrophys.\ J.} {\bf 678} 154--67

\item[] Li Y, Haiman Z and Mac Low M-M 2007 {\it Astrophys.\ J.}
  {\bf 663} 61--70

\item[] L\"ockmann U and Baumgardt H 2009 {\it Mon.\ Not.\ R. Astron.\
  Soc.} {\bf 394} 1841--6

\item[] Lodato G 2008 {\it New Astronomy Reviews} {\bf 52} 21--41

\item[] Maccarone T J and Servillat M 2008 {\it Mon.\ Not.\ R. Astron.\
  Soc.} {\bf 389} 379--84

\item[] Maciejewski W, Teuben P J, Sparke L S and Stone J M 2002, {\it
  Mon.\ Not.\ R. Astron.\ Soc.} {\bf 329} 502--12

\item[] Machida M N, Inutsuka S-I and Matsumoto T 2007 {\it Astrophys.\
  J.} {\bf 670} 1198--213

\item[] Magorrian J, Tremaine S, Richstone D, Bender R, Bower G, Dressler
  A, Faber S M, Gebhardt K, Green R, Grillmair C, Kormendy J and Lauer T
  1998 {\it Astron.\ J.} {\bf 115} 2285--305

\item[] Malbon R K, Baugh C M, Frenk C S and Lacey C G 2007 {\it Mon.\
  Not.\ R. Astron.\ Soc.} {\bf 382} 1394--414

\item[] Malmberg D and Davies M B 2009 {\it Mon.\ Not.\ R. Astron.\ Soc.}
  {\bf 394} L26--30

\item[] Mapelli M, Hayfield T, Mayer L and Wadsley J 2009 {\it Mon.\
  Not.\ R. Astron.\ Soc.} in press (astro-ph/0805.0185)

\item[] Maschberger Th and Clarke C J 2008 {\it Mon.\ Not.\ R. Astron.\
  Soc.} {\bf 391} 711--7

\item[] Mason B D, Hartkopf W I, Gies D R, Henry T J and Helsel J W 2009
  {\it Astron.\ J.} {\bf 137} 3358--77

\item[] Mathieu R D 1994 {\it Annu.\ Rev.\ Astron.\ Astrophys.} {\bf 32}
  465--530

\item[] Matsumoto T, Hanawa T and Nakamura F 1997 {\it Astrophys.\ J.}
  {\bf 478} 569--84

\item[] Mayer L, Kazantzidis S and Escala A 2009 {\it Mem.\ Soc.\
  Astron.\ Italiana} in press (astro-ph/0807.3329)

\item[] Mayor M, Udry S, Halbwachs J-L and Arenou F 2001 {\it The
  Formation of Binary Stars (IAU Symp.\ No.\ 200)} ed H Zinnecker
  and R D Mathieu (San Francisco: Astronomical Society of the Pacific)
  pp 45--54

\item[] Mazeh T, Simon M, Prato L, Markus B and Zucker S 2003 {\it
  Astrophys.\ J.} {\bf 599} 1344--56

\item[] McKee C F, Zweibel E G, Goodman ANA and Heiles C 1993 {\it
  Protostars and Planets III} ed E H Levy and J I Lunine  (Tucson:
  University of Arizona Press) pp 327--66

\item[] Mestel L 1965 {\it Quart.\ J. R. Astr.\ Soc.} {\bf 6} 161--98

\item[] Morris M and Serabyn E 1996 {\it Annu.\ Rev.\ Astron.\ Astrophys.}
  {\bf 34} 645--701

\item[] Mouschovias T Ch 1977 {\it Astrophys.\ J.} {\bf 211} 147--51

\item[] Mouschovias T Ch 1991 {\it The Physics of Star Formation and
  Early Stellar Evolution} ed C L Lada and N D Kylafis (Dordrecht:
  Kluwer) pp 61--122

\item[] Mouschovias T Ch and Ciolek G E 1999 {\it The Origin of Stars and
  Planetary Systems} ed C J Lada and N D Kylafis (Dordrecht: Kluwer)
  pp 305--39

\item[] Myers P C, Dame T M, Thaddeus P, Cohen R S, Silverberg R F, Dwek
  E and Hauser M G 1986 {\it Astrophys.\ J.} {\bf 301} 398--422

\item[] Myers P C, Evans N J and Ohashi N 2000 {\it Protostars and
  Planets IV} ed V Mannings \etal (Tucson: University of Arizona Press)
  pp 217--45

\item[] Nakano T, Nishi R and Umebayashi T 2002 {\it Astrophys.\ J.}
  {\bf 573} 199--214

\item[] Namekata D, Habe A, Matsui H and Saitoh T R 2009 {\it Astrophys.\
  J.} {\bf 691} 1525--39

\item[] Narita S, Hayashi C and Miyama S M 1984 {\it Prog.\ Theor.\
  Phys.} {\bf 72} 1118--36

\item[] Nayakshin S and King A 2009 {\it Mon.\ Not.\ R. Astron.\ Soc.} in
  press (astro-ph/0705.1686)

\item[] Nayakshin S, Cuadra J and Springel V 2007 {\it Mon.\ Not.\ R.
  Astron.\ Soc.} {\bf 379} 21--33

\item[] Norman M L, Wilson J R and Barton R T 1980 {\it Astrophys.\ J.}
  {\bf 239} 968--81

\item[] Ohashi N 1999 {\it Star Formation 1999} ed T Nakamoto (Nobeyama:
  Nobeyama Radio Observatory) pp 129--35

\item[] Ohashi N, Hayashi M, Ho P T P, Momose M, Tamura M, Hirano N and
  Sargent A I 1997 {\it Astrophys.\ J.} {\bf 488} 317--29

\item[] Ostriker E C 1994 {\it Astrophys.\ J.} {\bf 424} 292--318

\item[] Paumard T, Genzel R, Martins F, Nayakshin S, Beloborodov A M,
  Levin Y, Trippe S, Eisenhauer F, Ott T, Gillessen S, Abuter R, Cuadra
  J, Alexander T and Sternberg A 2006 {\it Astrophys.\ J.} {\bf 643}
  1011--35

\item[] Peiris H V and Tremaine S 2003 {\it Astrophys.\ J.} {\bf 599}
  237--57

\item[] Pfalzner S and Olczak C 2007 {\it Astron.\ Astrophys.} {\bf 462}
  193--8

\item[] Pfalzner S, Tackenburg J and Steinhausen M 2008 {\it Astron.\
  Astrophys.} {\bf 487} L45--8

\item[] Phinney E S 1994 {\it Mass-Transfer Induced Activity in Galaxies}
  ed I Shlosman (Cambridge: Cambridge University Press) pp 1--22

\item[] Preibisch T, Weigelt G and Zinnecker H 2001 {\it The Formation of
  Binary Stars (IAU Symp.\ No.\ 200)} ed H Zinnecker and R D Mathieu
  (San Francisco: Astronomical Society of the Pacific) pp 69--78

\item[] Price D J and Bate M R 2007 {\it Mon.\ Not.\ R. Astron.\ Soc.}
  {\bf 377} 77--90

\item[] Pudritz R E, Ouyed R, Fendt C and Brandenburg A 2007 {\it
  Protostars and Planets V} ed B Reipurth \etal (Tucson:
  University of Arizona Press) pp 277--94 

\item[] Ray T, Dougados C, Bacciotti F, Eisl\"offel J and Chrysostomou A
  2007 {\it Protostars and Planets V} ed B Reipurth \etal (Tucson:
  University of Arizona Press) pp 231--44

\item[] Reichard T A, Heckman T M, Rudnick G, Brinchmann J, Kauffmann G
  and Wild V 2009 {\it Astrophys.\ J.} {\bf 691} 1005--20

\item[] Reipurth B 2000 {\it Astron.\ J.} {\bf 120} 3177--91

\item[] Reipurth B 2001 {\it The Formation of Binary Stars (IAU Symp.\
  No.\ 200)} ed H Zinnecker and R D Mathieu (San Francisco: Astronomical
  Society of the Pacific) pp 249--60

\item[] Riechers D A, Walter F, Carilli C L and Lewis G F 2009 {\it
  Astrophys.\ J.} {\bf 690} 463--85

\item[] Robertson B, Hernquist L, Cox T J, Di Matteo T, Hopkins P F,
  Martini P and Springel V 2006 {\it Astrophys.\ J.} {\bf 641} 90--102

\item[] Romanova M M, Kulkarni A K and Lovelace R V E 2008 {\it
  Astrophys.\ J.} {\bf 673} L171--4

\item[] Rossa J, van der Marel R P, B\"oker T, Gerssen J, Ho L C, Rix
  H-W, Shields J C and Walcher C-J 2006 {\it Astron.\ J.} {\bf 132}
  1074--99

\item[] Sana H, Gosset E, Naz\'e Y, Rauw G and Linder N 2008 {\it Mon.\
  Not.\ R. Astron.\ Soc.} {\bf 386} 447--60

\item[] Sch\"odel R, Eckart A, Alexander T, Merritt D, Genzel R,
  Sternberg A, Meyer L, Kul F, Moultaka J, Ott T and Straubmeier C 2007
  {\it Astron.\ Astrophys.} {\bf 469} 125--46

\item[] Sellwood J A and Shen J 2004 {\it Coevolution of Black Holes and
  Galaxies} ed L C Ho (Cambridge: Cambridge University Press) pp 203--18 

\item[] Seth A, Ag\"ueros M, Lee D and Basu-Zych A 2008 {\it Astrophys.\
  J.} {\bf 678} 116--30

\item[] Shakura N I and Sunyaev R A 1973 {\it Astron.\ Astrophys.}
  {\bf 24} 337--55

\item[] Shen J and Sellwood J A 2004 {\it Astrophys.\ J.} {\bf 604}
  614--31

\item[] Shlosman I 1992 {\it Relationships Between Active Galactic Nuclei
  and Starburst Galaxies (ASP Conf.\ Ser.\ Vol.\ 31)} ed A V Filippenko
  (San Francisco: Astronomical Society of the Pacific) pp 335--46

\item[] Shlosman I and Begelman M C 1989 {\it Astrophys.\ J.} {\bf 341}
  685--91

\item[] Shlosman I, Frank J and Begelman M C 1989 {\it Nature} {\bf 338}
  45--7

\item[] Shlosman I, Begelman M C and Frank J 1990 {\it Nature} {\bf 345}
  679--86

\item[] Shu F H, Adams F C and Lizano S 1987 {\it Annu.\ Rev.\ Astron.\
  Astrophys.} {\bf 25} 23--81

\item[] Shu F, Najita J, Galli D, Ostriker E and Lizano S 1993 {\it
  Protostars and Planets III} ed E H Levy and J I Lunine (Tucson:
  University of Arizona Press) pp 3--45

\item[] Shu F H, Allen A, Shang H, Ostriker E C and Li Z-Y 1999 {\it The
  Origin of Stars and Planetary Systems} ed C J Lada and N D Kylafis
  (Dordrecht: Kluwer) pp 193--226

\item[] Shu F H, Najita J R, Shang H and Li Z-Y 2000 {\it Protostars and
  Planets IV} ed V Mannings \etal (Tucson: University of Arizona Press)
  pp 789--813

\item[] Simon M, Ghez A M, Leinert Ch, Cassar L, Chen W P, Howell R R,
  Jameson R F, Matthews K, Neugebauer G and Richichi A 1995 {\it
  Astrophys.\ J.} {\bf 443} 625--37

\item[] Somerville R S, Hopkins P F, Cox T J, Robertson B E and
  Hernquist L 2008 {\it Mon.\ Not.\ R. Astron.\ Soc.} {\bf 391} 481--506

\item[] Spitzer L 1968 {\it Nebulae and Interstellar Matter} ed B M
  Middlehurst and L H Aller (Chicago: University of Chicago Press)
  pp 1--63

\item[] Spitzer L 1978 {\it Physical Processes in the Interstellar
  Medium} (New York: Wiley-Interscience)

\item[] Stone J M, Gammie C F, Balbus S A and Hawley J F 2000 {\it
  Protostars and Planets IV} ed V. Mannings \etal (Tucson: University of
  Arizona Press) pp 589-611

\item[] Tan J C and Blackman E G 2005 {\it Mon.\ Not.\ R. Astron.\ Soc.}
  {\bf 362} 983--94

\item[] Tassis K and Mouschovias T Ch 2007 {\it Astrophys.\ J.} {\bf 660}
  388--401

\item[] Tomisaka K 2002 {\it Astrophys.\ J.} {\bf 575} 306--26

\item[] Toomre A 1964 {\it Astrophys.\ J.} {\bf 139} 1217--38

\item[] Trimble V 1990 {\it Mon.\ Not.\ R. Astron.\ Soc.} {\bf 242} 79--87

\item[] van de Ven G and Fathi K 2009 {\it Astrophys.\ J.} in press
  (astro-ph/0905.3556)

\item[] Vorobyov E I and Basu S 2006 {\it Astrophys.\ J.} {\bf 650}
  956--69

\item[] Wada K 2004 {\it Coevolution of Black Holes and Galaxies} ed
  L C Ho (Cambridge: Cambridge University Press) pp 186--201

\item[] Wada K and Norman C A 2002 {\it Astrophys.\ J.} {\bf 566} L21--4

\item[] Wardle M and Yusef-Zadeh F 2008 {\it Astrophys.\ J.} {\bf 683}
  L37--40

\item[] Watabe Y, Kawakatu N and Imanishi M 2008 {\it Astrophys.\ J.}
  {\bf 677} 895--905

\item[] Wehner E H and Harris W E 2006 {\it Astrophys.\ J.} {\bf 644}
  L17--20

\item[] Weidner C and Kroupa P 2006 {\it Mon.\ Not.\ R. Astron.\ Soc.}
  {\bf 365} 1333--47

\item[] Younger J D, Hopkins P F, Cox T J and Hernquist L 2008 {\it
  Astrophys.\ J.} {\bf 686} 815--28

\item[] Zinnecker H 1982 {\it Symposium on the Orion Nebula to Honor
  Henry Draper} ed A E Glassgold \etal {\it Ann.\ New York Acad.\ Sci.}
  {\bf 395} 226--35

\item[] Zinnecker H and Mathieu R 2001 (eds) {\it The Formation of Binary
  Stars (IAU Symp.\ No.\ 200)} (San Francisco: Astronomical Society of
  the Pacific)

\item[] Zinnecker H and Yorke H W 2007 {\it Annu.\ Rev.\ Astron.\
  Astrophys.} {\bf 45} 481--563

\endrefs

\end{document}